\documentclass[%
 aip,
 amsmath,amssymb,
 reprint,%
]{revtex4-1}

\usepackage{graphicx}
\usepackage{dcolumn}
\usepackage{bm}

\usepackage[utf8]{inputenc}
\usepackage[T1]{fontenc}
\usepackage{mathptmx}
\usepackage{etoolbox}
\usepackage{hyperref}
\usepackage{xcolor}

\makeatletter
\def\@email#1#2{%
 \endgroup
 \patchcmd{\titleblock@produce}
  {\frontmatter@RRAPformat}
  {\frontmatter@RRAPformat{\produce@RRAP{*#1\href{mailto:#2}{#2}}}\frontmatter@RRAPformat}
  {}{}
}%
\makeatother
\begin{document}

\preprint{AIP/123-QED}
\title[]{Terahertz magneto-nanoscopy of encapsulated monolayer graphene}
\author{Richard H. J. Kim}
\altaffiliation{Author to whom correspondence should be addressed: rkim@ameslab.gov, rkim@iastate.edu}
\affiliation{Ames National Laboratory, US Department of Energy, Ames, IA 50011, USA}%
\author{Sunwoong Yang}
\affiliation{Division of Science Education, Kangwon National University, Chuncheon 24341, Korea}%
\author{Taehoon Kim}
\affiliation{Department of Physics, Incheon National University, Incheon 22012, Korea}%
\author{Samuel J. Haeuser}%
\affiliation{Department of Physics and Astronomy, Iowa State University, Ames, IA 50011, USA}%
\author{Joong-Mok Park}
\affiliation{Ames National Laboratory, US Department of Energy, Ames, IA 50011, USA}%
\author{Randall K. Chan}%
\affiliation{Department of Physics and Astronomy, Iowa State University, Ames, IA 50011, USA}%
\author{Thomas Koschny}
\affiliation{Ames National Laboratory, US Department of Energy, Ames, IA 50011, USA}%
\author{Young-Mi Bahk}
\affiliation{Department of Physics, Incheon National University, Incheon 22012, Korea}%
\author{Sung Ju Hong}
\affiliation{Division of Science Education, Kangwon National University, Chuncheon 24341, Korea}%
\author{Jigang Wang}
\altaffiliation{jgwang@ameslab.gov, jgwang@iastate.edu}
\affiliation{Ames National Laboratory, US Department of Energy, Ames, IA 50011, USA}%
\affiliation{Department of Physics and Astronomy, Iowa State University, Ames, IA 50011, USA}%



\date{\today}

\begin{abstract}
This study investigates the nanoscale conductivity of encapsulated monolayer graphene at temperatures down to 5 K and magnetic fields of up to 1 T. We use the scattering-type scanning near-field optical microscopy (s-SNOM) technique to probe magnetic-field-dependent responses from graphene close to charge neutrality in the terahertz spectral region. We observe the near-perfect high-$q$ reflector behavior of graphene but with subtle changes by the presence of magnetic fields. Measurements align with calculations of the magneto-optical conductivity and the near-field spectroscopic contrast that describes the field-tunable cyclotron resonance of Dirac fermions. Our result provides an initial step toward understanding temperature and magnetic-field effects on nanoscale terahertz transport in two-dimensional quantum materials. 
\end{abstract}

\maketitle

\section{\label{sec:level1}Introduction}
Graphene has been the quintessential material for scanning near-field probes to visualize and investigate light-matter interactions happening at the nanoscale. The atomically thin crystalline layer offers a tip-scattered light detection virtually free of topographical crosstalk or edge artifacts, and exfoliated pristine samples of micron-scale sizes perfectly matches the field of view of conventional scanning probe microscopy stages. Using scattering-type scanning near-field optical microscopy (s-SNOM), propagating plasmons and their electrical tunability were first directly resolved in real space as interference fringes near the edges of graphene,\cite{chen,fei} and the research has started a worldwide interest in investigating polaritons from a variety of photonic systems. Extensions of the s-SNOM technique to extreme environments further enabled investigation on the fundamental limits of plasmon damping at low temperatures \cite{ni} and observation of magnetic-field-tunable magnetoexciton polaritons in graphene.\cite{dapo} While s-SNOM works have been domianted by studies conducted in the midinfrared spectral region, probing graphene in the farinfrared terahertz frequency region have soon followed up in recent years.\cite{apr,nrm} The near-field terahertz response of graphene was shown to exhibit high reflectivity comparable to noble metal films.\cite{zhang} Propagation of terahertz plasmons have been resolved in a highly confined environment, \cite{lund} and time-domain visualizations were realized utilizing spacetime imaging or field-resolved detection of tip-scattered ultrafast pulses.\cite{xu,angl} Meanwhile, many emergent quantum phenomena driven by electronic correlations and topology have been recently discovered in graphene-based systems, such as superconductivity in the magic angle twisted bilayer graphene \cite{cao} and fractional quantum anomalous Hall effect in a rhombohedral pentalayer graphene.\cite{lu} Optical properties of similar systems have been studied by infrared nanoimaging that has led to finding propagating plasmon modes in moir\'{e} patterned twisted bilayer graphene \cite{sunku,hesp} and identifying layer-stacking domains and domain walls in bilayer and tetralayer graphene.\cite{ju,wirth} To this end, further exploration of nanoscale quantum phenomena will require a terahertz s-SNOM measurement incorporating cryogenic operation and applied magnetic fields that can probe local conductivity and low-energy excitations in these graphitic materials.

Here, we provide a convenient starting point toward this direction by studying single layer graphene at low temperatures down to 5 K and at applied magnetic fields of up to 1 T. We employ s-SNOM to probe graphene conductivity in the deep subwavelength scale at energies in the terahertz frequency range. Our measurements of the scattered near-field signal as a function of frequency and magnetic field agrees qualitatively with the calculations based on the magneto-optical conductivity of graphene that depends on the formation of quantized Landau levels of Dirac fermions. The results pave the way for future in-depth assessment of local magnetotransport properties in two-dimensional systems at the nanoscale.

\section{Results and discussion}
\begin{figure} [!ht]
\includegraphics[width=\columnwidth]{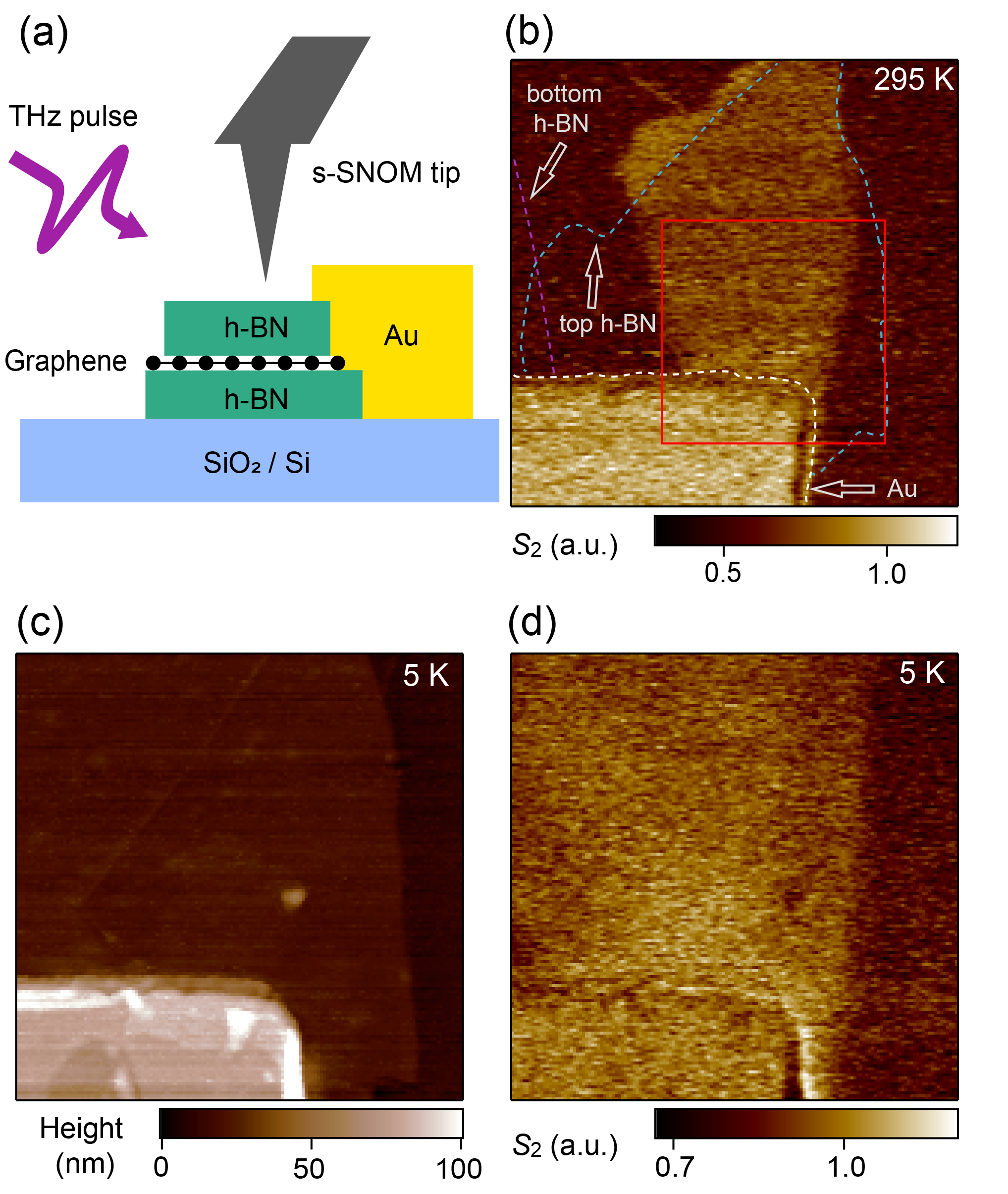}
\caption{Terahertz nanoimaging of encapsulated monolayer graphene. (a) Schematic of the monolayer graphene sandwiched by h-BN flakes and the terahertz near-field measurement scheme based on an atomic force microscope. (b) A 20 $\mu$m by 20 $\mu$m terahertz near-field image of graphene at room temperature. The borders of top and bottom h-BN and gold electrode are respectively outlined by blue, purple, and white dashed lines for clarity. Low temperature (c) topography and (d) terahertz near-field mapping of graphene in the red boxed region shown in (b) at zero magnetic field. Graphene appears bright in (b) and (d).
\label{fig1}}
\end{figure}

Figure \ref{fig1}(a) shows the schematic of our terahertz near-field measurement of graphene. Monolayer graphene is encapsulated between two flakes of h-BN. Both graphene and h-BN flakes were mechanically exfoliated on a SiO$_{2}$ (300 nm) / Si substrate using adhesive tape. A dry-transfer method utilizing a polycarbonate (PC) film was employed to assemble the h-BN/graphene/h-BN heterostructure.\cite{purd} First, a top h-BN flake with thickness of $\sim$10 nm was picked up by a PC stamp, and subsequently the graphene flake was picked up using the h-BN/PC stamp (90 $^{\circ}$C); the graphene layer was partially encapsulated by the top h-BN, leaving an exposed region for electrode contact. Second, the graphene/h-BN/PC stamp was transferred onto a bottom h-BN flake with thickness of $\sim$30 nm at 150 $^{\circ}$C, followed by the removal of the PC film in chloroform. Finally, the h-BN/graphene/h-BN heterostructure was annealed at 350 $^{\circ}$C and below 10$^{-5}$ torr for 3 hours. Using a standard photolithography and metallization process, the electrode with Cr (5 nm) / Au (75 nm) was fabricated and was used as a reference for near-field signal analysis. For s-SNOM, terahertz pulses are generated from a GaP crystal by using a ytterbium laser with a pulse energy of 20 $\mu$J, a repetition rate of 1 MHz, and a central laser wavelength of 1030 nm, and the near-field terahertz waveforms are detected by electro-optical sampling from a CdTe crystal. We use a cantilever-based tapping-mode scanning probe microscope system so that the near-field amplitude is extracted from the scattered terahertz signal by demodulating the backscattered radiation collected from the sample area in the vicinity of the s-SNOM tip (model 25Pt300B, Rocky Mountain Nanotechnology) at second harmonics of the tip-tapping frequency ($S_{2}$).\cite{n4,n6} A split pair magnet cryostat is used for low temperature measurements, and the superconducting magnet surrounding the sample space supplies a static magnetic field perpendicular to the sample surface.\cite{kim} 

To obtain near-field images, the sample stage underneath the tip is raster scanned, while the sampling delay to resolve the terahertz field is fixed to a position that gives the largest near-field signal amplitude. The terahertz near-field image of the graphene area taken at room temperature is displayed in Fig. \ref{fig1}(b). Graphene and the gold electrode appear bright, whereas h-BN and the substrate are shown to be indistinguishably dark in the image. The same behavior is observed at 5 K as shown in Fig. \ref{fig1}(c) and (d) that presents the simultaneously taken height and terahertz near-field map, respectively. Terahertz nanoimaging clearly reveals graphene while only the top h-BN and gold electrode can be outlined from the topography scan. The scattering amplitude from graphene is slightly less than gold likely due to the top h-BN preventing a close contact of graphene with the tip and the carrier concentration being minimal close to charge neutrality. At 5 K, the conductivity of graphene in the terahertz region still lies in the order of $e^{2}/h$ ($e$ is the elementary charge and $h$ is the Planck constant) either due to interband transitions if it is at the charge neutrality point or because of the dominating intraband contribution that already starts at a charge carrier concentration of $n_{\mathrm{c}}=1.0\times10^{9}$ cm$^{-2}$, raising the conductivity even higher. This doping level leads to a Fermi energy of $E_{\mathrm{F}}=\hbar v_{\mathrm{F}}\sqrt{\pi n_{\mathrm{c}}}=4.1$ meV ($\hbar$ is the reduced Planck constant) which is equivalent to a photon energy at 1 THz, where $v_{\mathrm{F}}=1.1\times10^{6}$ m/s is the Fermi velocity. Thus, the high conductivity from the semimetallic terahertz response turns graphene into a one-atom-thick perfect reflector for high in-plane momentum. Also, no particular spatial inhomogeneities in conductivity have been observed within the noise level of our image scans with or without the application of magnetic fields.

\begin{figure} [!ht]
\includegraphics[width=\columnwidth]{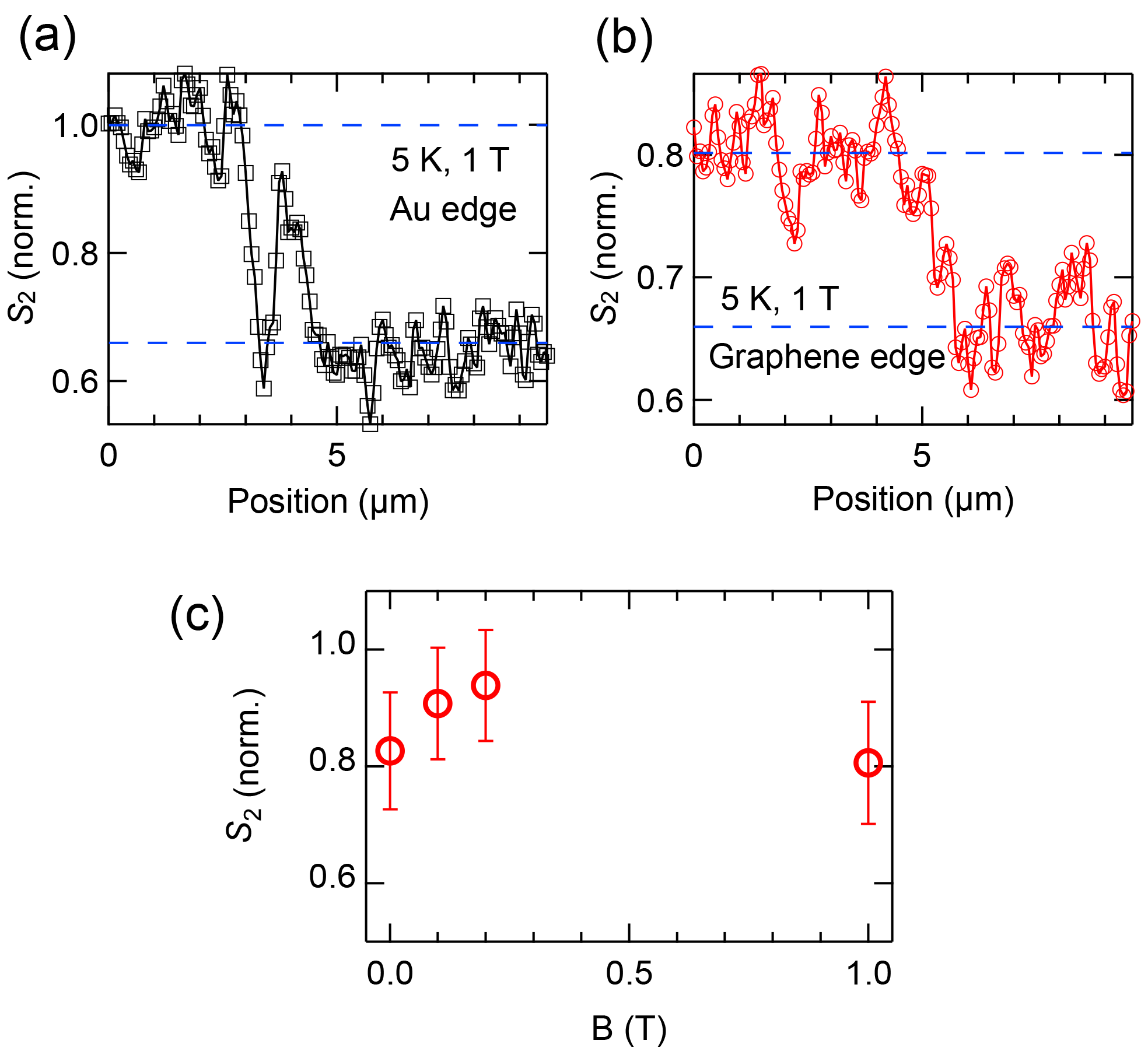}
\caption{Magnetic-field-dependent terahertz near-field contrast of graphene. Terahertz near-field amplitude line scans across (a) a gold microstructure edge and (b) the edge of graphene at 5 K and 1 T. Gold and graphene lies on the left half in each of the line scans. (c) Average and standard deviations of near-field signals from graphene line scans at different magnetic fields from 0 to 1 T. The scattered signals $S_{2}$ are normalized to the average value obtained from gold each taken at the corresponding field value.
\label{fig2}}
\end{figure}

Next, we performed a horizontal line scan across the gold electrode and the graphene edge as shown in Fig. \ref{fig2}(a) and (b), respectively. At a field of 1 T, graphene gives a scattered amplitude $S_{2}$ value of 0.80, and the area covered only by h-BN yields 0.66 when the signal from gold is normalized to 1. The same measurement was carried out at field strengths of 0, 0.1, and 0.2 T at the same temperature of 5 K, and the results of the normalized $S_{2}$ for graphene is plotted in Fig. \ref{fig2}(c). Overall, the signal from graphene stays more or less high in this field range. Although fluctuation from the noise level prevents a detailed measurement of the signal variation with field strength, a slight decrease in the signal at the highest 1 T may be plausible as can be inferred from the calculations which will be discussed later.

\begin{figure} [!ht]
\includegraphics[width=\columnwidth]{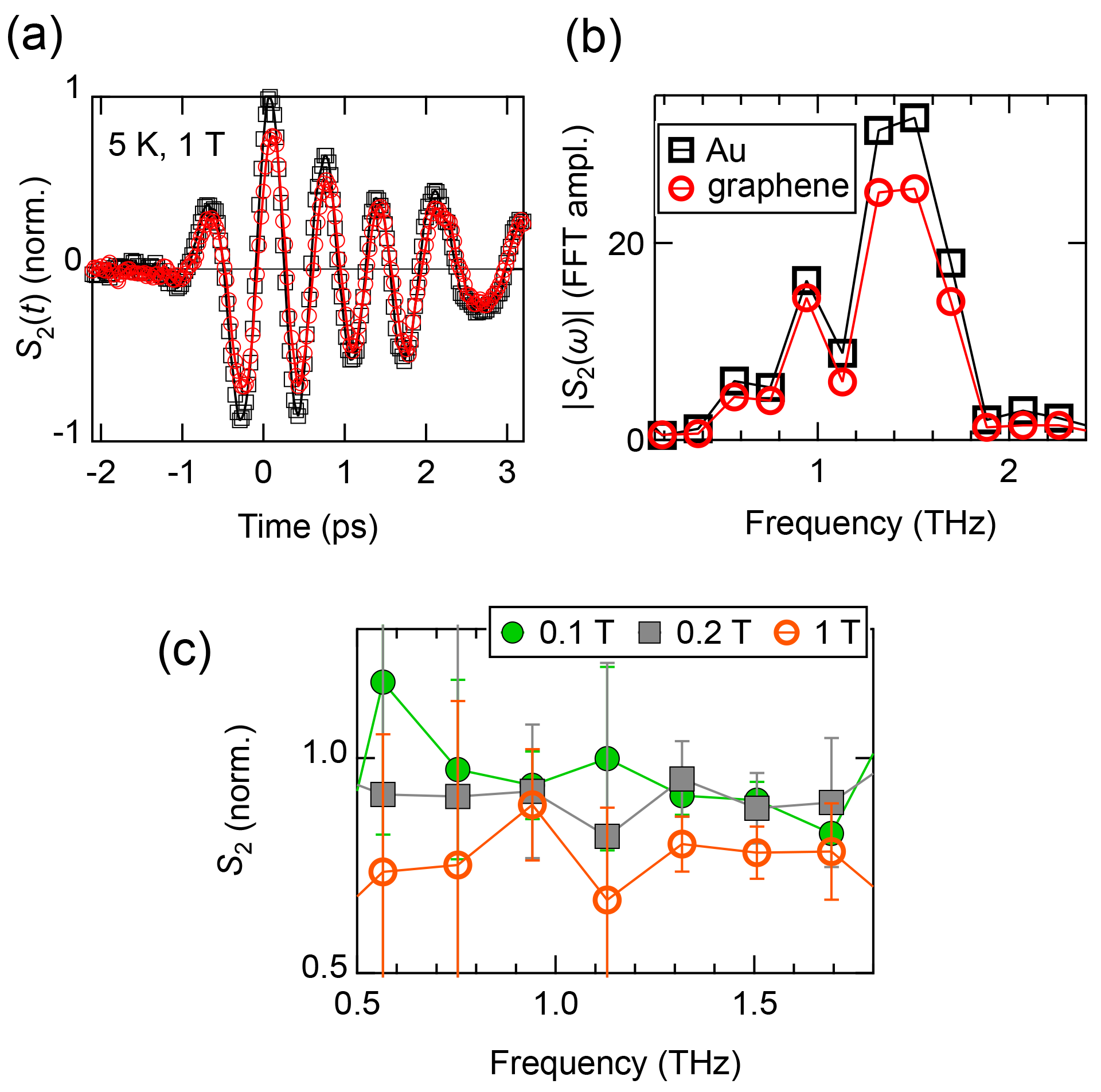}
\caption{Terahertz magneto-nanospectroscopy of graphene. (a) Terahertz near-field time traces measured on gold and graphene at 5 K and 1 T. (b) Fourier-transformed spectra of the time traces in (a). (c) Relative amplitude spectra of graphene with respect to the spectra taken from gold at fields of 0.1, 0.2, and 1 T. Standard deviations were calculated from data points showing the noise level at the beginning of the time-domain traces.
\label{fig3}}
\end{figure}

We also conducted a separate time-domain spectroscopy measurement with gold used for obtaining a reference spectra to derive the response from graphene. The time traces for gold and graphene taken at 1 T and their respective fast-Fourier-transformed spectra are displayed in Fig. \ref{fig3}(a) and (b), respectively. The relative amplitude spectrum of graphene from 0.5 to 1.8 THz is then plotted in Fig. \ref{fig3}(c) together with the spectra attained at 0.1 and 0.2 T. Interestingly, we again observe a slight depression in the measured values at 1 T in the measured near-field terahertz spectrum at 5 K.

\begin{figure} [!ht]
\includegraphics[width=\columnwidth]{fig4}
\caption{Magneto-optical conductivity and measured terahertz scattering near-field amplitude of graphene. Calculated (a) real-part and (b) imaginary part of the longitudinal conductivity of graphene at different terahertz frequencies as a function of the magnetic field. (c) Calculated $S_{2}$ amplitude of graphene with respect to gold as a function of magnetic field using the conductivity values from (a) and (b). Experimental values are identical to those shown in Fig. \ref{fig2}(c). (d) Calculated $S_{2}$ amplitude of graphene with respect to gold as a function of frequency for different magnetic field values and the experimental values taken from Fig. \ref{fig3}(c).
\label{fig4}}
\end{figure}

The magneto-optical conductivity of graphene is calculated to deduce the near-field terahertz response. In a magnetic field $B$, the longitudinal conductivity $\sigma_{xx}(\omega)$ of graphene at a photon energy of $\hbar\omega$ is expressed as \cite{dapo,gusy}
\begin{eqnarray}
&&\sigma_{xx}(\omega)=i\frac{e^{2}v_{\mathrm{F}}^{2}|eB|(\hbar\omega+i\Gamma)}{\pi}\times\nonumber\\
&&\sum_{n=0}^{\infty}\biggl\{\frac{1}{E_{n}-E_{n+1}}\frac{f(E_{n})-f(E_{n+1})+f(E_{-(n+1)})-f(E_{-n})}{(E_{n}-E_{n+1})^{2}-(\hbar\omega+i\Gamma)^{2}}\nonumber\\
&&+\frac{1}{E_{-n}-E_{n+1}}\frac{f(E_{-n})-f(E_{n+1})+f(E_{-(n+1)})-f(E_{n})}{(E_{-n}-E_{n+1})^{2}-(\hbar\omega+i\Gamma)^{2}}\biggr\},\nonumber\\
\label{eq1}
\end{eqnarray}
where $f(E_{n})=1/e^{(E_{n}-E_{\mathrm{F}})/k_{\mathrm{B}}T}+1$ is the Fermi-Dirac distribution function, $E_{n}=\mathrm{sgn}(n)\sqrt{2\hbar|neB|v_{\mathrm{F}}^{2}}$ is the energy with Landau level index $n$, and we assume a phenomenological width of $\Gamma=8.0$ meV associated with the Landau level transitions \cite{dapo} which also gives a reasonable electron scattering time of $\hbar/\Gamma=82$ fs in the $B\rightarrow0$ limit. The real and imaginary part of the conductivity as a function of the magnetic field for different terahertz frequencies are shown in Fig. \ref{fig4}(a) and (b), respectively. Here we assume a charge neutral graphene, and the sum of the Landau level transitions is taken up to a suitable large number of $n=100,000$ in Eq. (\ref{eq1}). Note that an infinite number of transitions need to be taken into account in the limit of $B\rightarrow0$. 

After calculating the conductivity, the finite dipole model \cite{cvit,agha,lass} and a transfer matrix model for graphene \cite{lass,zhan,wirt} was used to estimate the terahertz near-field contrast. From the finite dipole model, the tip-scattered field is given by $\tilde{E}_{\mathrm{sca}}\propto(1+r_{\mathrm{p}}^{2})\alpha_{\mathrm{eff}}\tilde{E}_{\mathrm{inc}}$, where $r_{\mathrm{p}}$ is the far-field Fresnel reflection coefficient of the sample for $p$-polarized light, $\tilde{E}_{\mathrm{inc}}$ is the incident electric field, and $\alpha_{\mathrm{eff}}$ is the effective polarizability of the tip-sample system. Here $\alpha_{\mathrm{eff}}=C(1+f_{0}\beta/2(1-f_{1}\beta))$, where $f_{0}=(0.7-\frac{2H+W_{0}+a}{2L})\frac{\mathrm{ln}(4L/(4H+2W_{0}+a))}{\mathrm{ln}(4L/a)}$ and $f_{1}=(0.7-\frac{2H+W_{i}+a}{2L})\frac{\mathrm{ln}(4L/(4H+2a))}{\mathrm{ln}(4L/a)}$ with the effective tip length $L=600$ nm, tip radius $a=40$ nm, $W_{0}=1.31aL/(L+2a)$, $W_{i}=a/2$, $C=La^{2}$, tip--sample distance $H(t)=\frac{1}{2}A(1+\mathrm{sin}\Omega t)$, $A=200$ nm, and tip tapping frequency $\Omega=14.49$ kHz. 

We then utilize the matrix-based method to obtain the near-field reflection coefficient $\beta$ from the layered graphene sample that includes SiO$_{2}$, and the Si substrate. For simplicity, we omit the h-BN layers and observe how the Landau level transitions are translated into the terahertz optical response. Considering the propagation of light across a conductive interface with conductivity $\sigma$ that separates two dielectrics with dielectric constants $\epsilon_{1}$ and $\epsilon_{2}$, the transfer matrix for $p$-polarized light between the two layers is defined by $D_{12}=\frac{1}{2}\begin{bmatrix}
1+\eta_{12}+\xi_{12} & 1-\eta_{12}-\xi_{12}\\
1-\eta_{12}+\xi_{12} & 1+\eta_{12}-\xi_{12}
\end{bmatrix}$, with $\eta_{12}=\epsilon_{1}k_{2z}/\epsilon_{2}k_{1z}$, $\xi_{12}=\sigma k_{2z}/\epsilon_{0}\epsilon_{2}\omega$, $k_{1(2)z}=\sqrt{\epsilon_{1(2)}\omega^{2}/c^{2}-q^{2}}$, $q\approx1/a$, $\epsilon_{0}$ is the vacuum permittivity, and $c$ is the speed of light. Here we have $\sigma=\sigma_{xx}(\omega)$ between air and SiO$_{2}$, and $\sigma=0$ between SiO$_{2}$ and Si. Also, the propagation matrix for a medium with thickness $\Delta z$ is described by $P(\Delta z)=\begin{bmatrix}
e^{-ik_{z}\Delta z} & 0\\
0 & e^{ik_{z}\Delta z}
\end{bmatrix}$. The full transfer matrix of the total sample structure becomes $\mathcal{M}=D_{12}P(\Delta z)D_{23}$, where we use $\epsilon_{1}=1$ for air, $\epsilon_{2}=4$ \cite{fran} and $\Delta z=300$ nm for SiO$_{2}$, and $\epsilon_{3}=11.9$ for Si. It is then calculated that $\beta(\omega,q)=M_{21}/M_{11}$ from the elements of $\mathcal{M}$. We finally obtain the harmonic orders of the scattered signal by Fourier decomposition of the time-dependent signal. For the second harmonic $S_{2}$, we have $S_{2}\propto\int_{0}^{T}\tilde{E}_{\mathrm{sca}}(t)e^{i2\Omega t}dt$ where $T$ is the tapping period and the time dependence comes from the tip--sample distance $H(t)$. Using the conductivity values from Fig. \ref{fig4}(a) and (b), $S_{2}$ as a function of magnetic field is calculated as shown in Fig. \ref{fig4}(c). The $S_{2}$ values are normalized by the calculated $S_{2}$ of a 300-nm-thick gold film with the Drude parameters of plasma frequency $\omega_{p}=2\pi\times2183$ THz and scattering rate $\gamma=2\pi\times6.46$ THz.\cite{ordal} The principal cyclotron resonance from the relativistic Landau levels with $0\rightarrow1$ ($-1\rightarrow0$) transition becomes distinguishable in the near-field scattering signal when it is probed at frequencies higher than $\sim\Gamma$.

Also, a plot of $S_{2}$ as a function of frequency for different magnetic fields is shown in Fig. \ref{fig4}(d). For the curve at 0 T, we use the temperature ($T$)-dependent analytic expression for the graphene conductivity $\sigma_{xx}(\omega)$ evaluated in the local limit \cite{zhang,gusy,kopp}
\begin{eqnarray}
&&\sigma_{xx}(\omega; B\rightarrow0)=\frac{2k_{\mathrm{B}}Te^{2}}{\pi\hbar^{2}}\frac{i}{\omega+i\Gamma/\hbar}\mathrm{ln}\left[2\mathrm{cosh}\left(\frac{E_{\mathrm{F}}}{2k_{\mathrm{B}}T}\right)\right]\nonumber\\
&&+\frac{e^{2}}{4\hbar}\left[H\left(\frac{\omega}{2}\right)+i\frac{4\omega}{\pi}\int_{0}^{\infty}dx\frac{H(x)-H\left(\frac{\omega}{2}\right)}{\omega^{2}-4x^{2}}\right],
\label{eq2}
\end{eqnarray}
where
\[H(x)=\mathrm{sinh}(\hbar x/k_{\mathrm{B}}T)/[\mathrm{cosh}(E_{\mathrm{F}}/k_{\mathrm{B}}T)+\mathrm{cosh}(\hbar x/k_{\mathrm{B}}T)].\]
In Fig. \ref{fig4}(c) and Fig. \ref{fig4}(d), the data from Fig. \ref{fig2}(c) and Fig. \ref{fig3}(c) are plotted together, respectively. Both results indicate that the near-field amplitude gradually decreases with higher frequency and magnetic field if we do not probe near the cyclotron resonance. We further note that future experiments with bandwidths extending up to 3 to 10 THz can provide ways to observe these resonances at sub-1 T magnetic fields.

\begin{figure} [!ht]
\includegraphics[width=0.8\columnwidth]{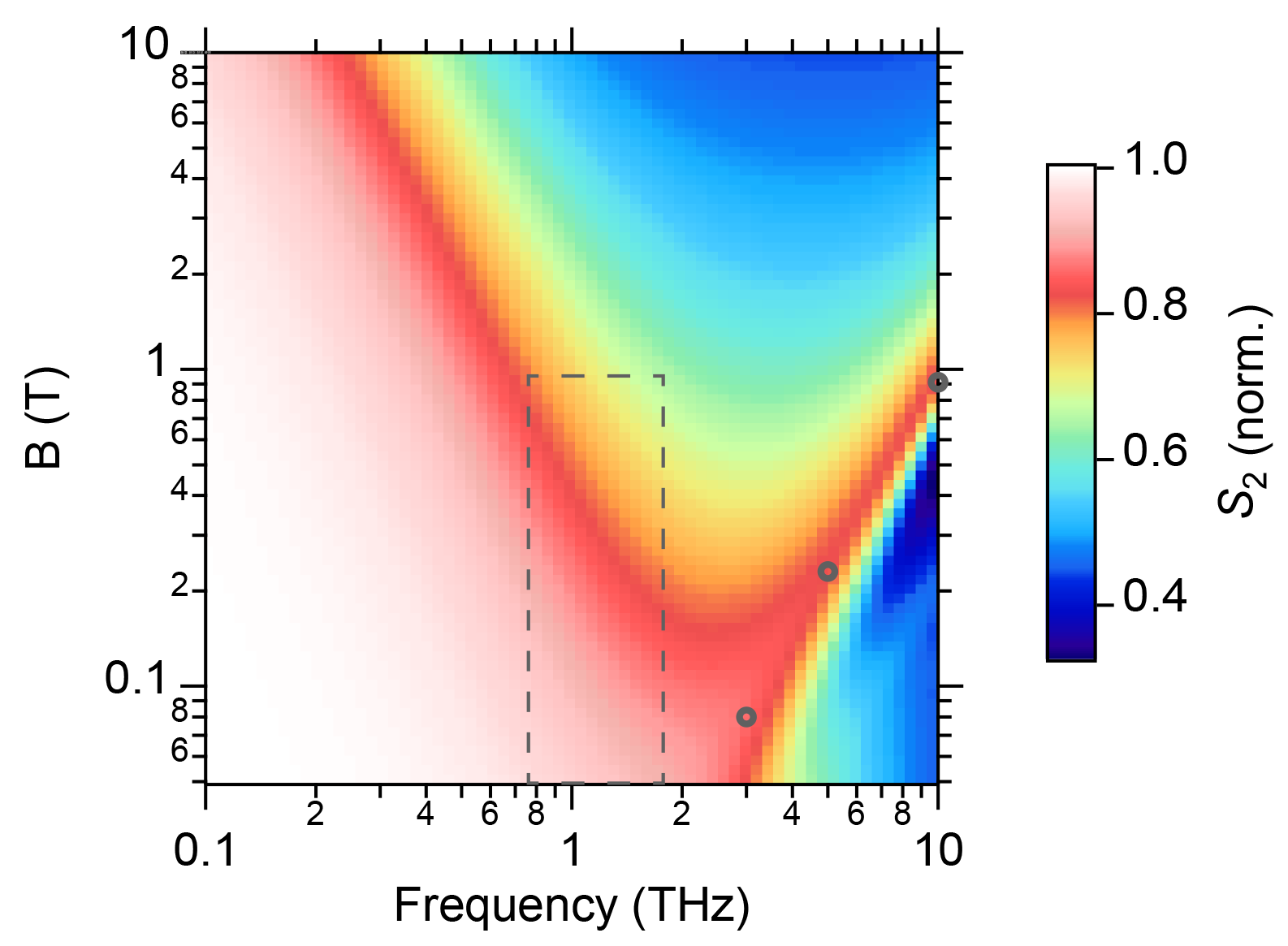}
\caption{Calculated $S_{2}$ of graphene as a function of magnetic field and frequency in the terahertz range using Eq. (\ref{eq1}) together with the finite dipole model and the transfer matrix model. The dashed box represents the space surveyed in this study, and circled markers correspond to the cyclotron resonance peaks at 3, 5, and 10 THz as presented in Fig. \ref{fig4}(c).
\label{fig5}}
\end{figure}

Lastly, the near-field scattering amplitude as a function of both $B$ and $\omega$ was calculated as displayed in Fig. \ref{fig5}. The plot shows that $S_{2}$ generally decreases with stronger fields and higher frequency if we exclude the peak from the Landau level transition. This follows from Eq. (\ref{eq1}) for the conductivity values at large $B$ or $\omega$. Also, as the dielectric constant that describes the Drude response is proportional to the conductivity divided by the optical frequency, the dielectric contrast becomes less at higher frequencies and leads to a reduced $S_{2}$, which can be recognized by the smaller $S_{2}$ values at $B=0$ for higher probing frequencies in Fig. \ref{fig4}(c) despite similar conductivity numbers here as can be extracted from Fig. \ref{fig4}(a) and (b). The two-dimensional plot indicates that our measurements of around 1 THz and in the range below 1 T belongs to a threshold region where $S_{2}$ falls below 90\% of the near-field signal expected from a perfect metal. Thus, graphene will no longer act as a good near-field reflector outside this region.

\section{Conclusion}
We measured the nanoscale terahertz response of graphene using s-SNOM at 5 K temperatures and fields of up to 1 T. We observe the boundaries of the field amplitude and probing frequency that define the $B$ and $\omega$ space where charge-neutral graphene is valid as a near-perfect reflector for evanescent terahertz fields. The stability from moderate magnetic perturbations suggests potential benefits of graphene for device applications at terahertz frequencies. Our study expands nanoscale conductivity measurements of graphene to low temperatures and high fields and will open up opportunities to probe real-space variations of low-energy spectroscopic signatures of electron correlations and topology phenomena in two-dimensional crystals and other encapsulated superlattice devices.

\begin{acknowledgments}
This work was supported by the US Department of Energy, Office of Basic Energy Science, Division of Materials Sciences and Engineering (Ames National Laboratory is operated for the US Department of Energy by Iowa State University under contract no. DE-AC02-07CH11358). This work was partially supported by National Research Foundation of Korea (NRF) grants funded by the Korean government (NRF-2022R1A2C1011655 (Y.-M.B.) and RS-2025-16067067 (S.J.H.)). 
\end{acknowledgments}

\section*{Data Availability Statement}

The data that support the findings of this study are available within the article or are available from the corresponding author upon reasonable request.

\section*{Author Declarations}
\subsection*{Conflict of Interest}
The authors have no conflicts to disclose.

\nocite{*}
\bibliography{aipsamp}

\end{document}